\documentstyle[11pt,a4,psfig]{article}
\begin{document}
\rightline{IMPERIAL/TP/94-95/01}
\rightline{NI94018}

\vskip 3cm

\begin{center}
{\LARGE High-Harmonic Configurations of Cosmic Strings:\\
        An Analysis of Self-Intersections}
        \end{center}

        \begin{center}
        Xavier A. Siemens$^a$ and T. W. B. Kibble$^{a,b}$\\
        {\small$^a$Blackett Laboratory, Imperial College, London
SW7 2BZ, UK}\\
        {\small$^b$Isaac Newton Institute for Mathematical Sciences,}\\
        {\small 20 Clarkson Road, Cambridge CB3 0EH, UK}
        \end{center}

\baselineskip 1.2 em

\begin{abstract}
A general formulation for describing odd-harmonic cosmic strings is
developed and used to determine the self-intersection properties of
high-harmonic loops. This is important because loop formation mechanisms
produce high-harmonic components (kinks) which can only be eliminated
very slowly by gravitational radiation, damping by the dense
surrounding plasma in the era of string formation, or by the expansion of
the Universe. For the class of loops examined it has been found that in
the high-harmonic limit, essentially all cosmic loops self-intersect.
\end{abstract}

\section{Introduction}
Topological defects such as monopoles, cosmic strings and textures,
are predicted by a number of grand unified theories as a result of
symmetry breaking and may have important comological consequences
\cite{Kib76}. Cosmic strings, in particular, could be responsible for the
creation of the density fluctuations needed for large-scale structure
formation \cite{ZelVil81,dens}. They are characterized by a
tension or mass per unit length $\mu$ which is related to the
temperature $T_{\rm GUT}$ at which the phase transition takes
place by $\mu \sim T_{\rm GUT}^2$. Cosmic strings can meet and
intercommute (exchange partners) or self-intersect. The latter process
results in the formation of cosmic loops.

In the old cosmic string scenario \cite{AT85,old,Ben86} it was envisaged
that these loops gave rise to the density fluctuations responsible for
galaxy, galaxy cluster  and supercluster formation. Loops decay
primarily via gravitational radiation, giving them a very long lifetime.
Recent numerical simulations of cosmic string networks
\cite{AT85,AT89,BB88,BB89,AS}, while confirming that a scaling solution is
probably reached, have shown that there is a great deal of small-scale
structure on the strings, and that the typical size of the loops formed
is very small compared to the horizon.  Since the lifetime of a loop is
proportional to its length, this means that they do not live very long,
and are unable to act as the seeds around which galaxies form.  The
dominant density perturbations must come from the long strings.
Nevertheless the loops could still have significant effects
collectively.  The extent to which they do so depends strongly on
whether they typically break up into large numbers of small loops or
reach a stable form that can survive a long time.

It is difficult to analyze the fate of such loops using the numerical
simulations of cosmic string networks, because they lack the necessary
resolution.  Bennett and Bouchet \cite{BB88} for example estimated that
most loops break up into about ten stable daughter loops, but since the
resulting stable loops are comparable in size to the resolution, it is
unclear how far we should trust this conclusion.  So it is desirable to
study the evolution of loops analytically.

Several string parametrizations, corresponding to low harmonics (first
and third), have been analyzed for self-intersections --- those of
Kibble and Turok \cite{KT82}, a one-parameter family of cosmic loops,
Turok \cite{Tur84}, with two parameters, Chen, DiCarlo and Hotes
\cite{CDH}, with three and finally DeLaney, Engle and Scheick \cite{DES},
with five.  The principal justification for limiting study to such
configurations (aside from computational convenience) is the argument
that high-harmonic components are damped either by the
expansion of the Universe, by friction with the dense surrounding plasma
in the era immediately after string formation, or by the gravitational
back-reaction, so that after a time loops become relatively smooth
\cite{Tur84,Ben86,CDH,DES}.  However, we shall argue that these mechanisms
may not in fact be so relevant to loop stability.

\begin{figure}[hbt]
  \begin{center}
   \begin{tabular}{ccc}
   \end{tabular}
  \end{center}
\caption{String intercommutation (a) and loop formation mechanisms:
By the self-intersection of a long string (b) and by
the self-intersection of a cosmic loop (c).}
\label{fig:intersect}
\end{figure}

Loop formation mechanisms are illustrated in Figure 1.  When strings meet
they intercommute (Fig.\ 1(a)). Cosmic loops can be
created by the self-intersection of a pre-existing loop (Fig.\ 1(b)) or
of a long string (Fig.\ 1(b)).  Both mechanisms inevitably
produce kinks, or equivalently, high-harmonic components.  Damping by
interaction with the dense surrounding plasma only affects strings
shortly after their initial formation.  However, the loops of
cosmological interest are produced when these effects are no longer
important.  Similarly, the expansion of the Universe is relevant only
for loops comparable in size to the horizon, not to the small loops
that dominate the recent loop distribution.   High harmonics can also be
radiated away by the emission of gravitational waves.  The time scale of
gravitational radiation effects, however, is very much larger than the
period of oscillation of a cosmic loop. Therefore, since a cosmic loop in
a self-intersecting trajectory will self-intersect within one oscillation
period, it is reasonable to consider gravitational effects negligible as
far as loop stability is concerned.

For all these reasons, it would seem therefore to be useful to
investigate the properties of higher-harmonic cosmic loops to determine
how the self-intersection probability varies with the number of
harmonics. This is the problem we address in this work.  The presence of
these extra harmonics makes the parameter space of cosmic loops larger
and more complex than that of the string configurations analyzed so far.

Section 2 provides a summary of the string equations of motion and the
general solutions in terms of Fourier series and in terms of the product
of rotations \cite{BD91}. Section 3 develops a general formulation for an
odd-harmonic cosmic string in terms of the product representation, and
describes our method for finding self-intersections.  An
explanation of how it is computationally implemented is given in an
Appendix.  In section 4 we describe the self-intersection results
obtained for harmonic parametrizations which range from a $3/3$ string
({\it i.e.}, one including up to the third harmonic on each of the
left- and right-moving halves) up to a $41/41$ string.  We also discuss
their sensitivity to different cutoffs. Finally, Section 5 contains a
discussion of the new self-intersection results, their effect on
galaxy-formation scenarios, and new directions of research.

\section{Equations of Motion and General Solutions}

The position of a string is described by $x^\mu(\sigma,\tau)$, where
$\sigma$ and $\tau$ are the spacelike and timelike variables
parametrizing the world sheet that the string sweeps out in space-time
($c$ is taken to be 1).  It is convenient to use the {\it orthonormal\/}
(or {\it conformal\/}) gauge defined by the constraints \cite{GGRT}
\begin{equation}
\partial _{\sigma} x ^{\mu} \partial _{\tau} x _{\mu} =
\partial _{\sigma} x ^{\mu} \partial _{\sigma} x _{\mu} +
\partial _{\tau} x ^{\mu} \partial _{\tau} x _{\mu}=0.
\end{equation}
For loops small compared with the horizon size and when
gravitational radiation and external gravitational fields are
considered negligible, the equation of motion of a cosmic
string is the classical relativistic wave equation
\begin{equation}
\partial ^2 _{\sigma} x^{\mu} - \partial ^2 _{\tau} x^{\mu}
= 0.
\end{equation}
We are also free to impose the further condition $x^0=t$, so that
$x ^{\mu} ( \sigma ,t) =(t,{\bf r} (\sigma ,t))$.  Then
the solution in the centre of mass frame can be written
\begin{equation}
{\bf r} (\sigma ,t)= {1 \over 2} [{\bf a} (u)+{\bf b} (v)],
\end{equation}
with
$$u= \sigma -t,\quad v= \sigma +t,$$
and the constraints become
\begin{equation}
\biggl[{d{\bf a} \over du}\biggr] ^2 =
\biggl[{d{\bf b} \over dv}\biggr] ^{2} = 1.
\label{constr}
\end{equation}

These vectors trace out closed loops on a unit sphere \cite{KT82}.
Here they will be referred to as the two halves of the string.

We can choose, for convenience, the period in $\sigma$ to be $2 \pi$.
Thus {\bf r} is periodic in $t$ with effective period $\pi$.
The two halves of a closed string loop can be expanded in
terms of the Fourier series
\begin{equation}
\begin{array}{rcl}
{\bf a}'&=& {\bf Z} + \sum  {\bf A }_n \cos(nu)+ \sum{\bf B}_n
\sin(nu),\\
{\bf b}'&=& -{\bf Z} + \sum  {\bf C }_n \cos(nv)+ \sum{\bf D}_n
\sin(nv),
\end{array}
\label{fourier}
\end{equation}
where ${\bf a}'$ and ${\bf b}'$ do not, in general, have
the same number of harmonics.
The linear centre of mass (c.m.)\ terms of ${\bf a}$ and ${\bf b}$
must be related due to spatial periodicity of closed loops.
The problem therefore is reduced to finding a set of harmonic
coefficients that satisfy the constraint equations (\ref{constr})
and the relation
between the c.m.\ terms of ${\bf a}$ and ${\bf b}$ (\ref{fourier}).

The general solution of an $N$-harmonic string half requires the
satisfaction of $4N+1$ nonlinear relations between the vector
coefficients. For the $N$=1, $N$=2 and $N$=3 cases these relations
can be solved with relative
ease but for higher harmonics they become intractable.
An alternative procedure is available involving
products of rotations \cite{BD91}. In terms of this representation a
general $N$-harmonic string half is given by
\begin{equation}
{\bf c}' _{N} (u)= \rho _{N+1} R_{z} (u) \rho _{N} R_{z} (u)
\ldots \rho _{3} R_{z} (u)  \rho _{2} R_{z} (u) \rho _{1} {\bf
{\hat z}}
\label{final}
\end{equation}
where
\begin{equation}
\rho _{i} = \rho ( \theta _{i},\phi _{i} ) =R_{z} (-
\theta_{i} )
R_{x} ( \phi _{i} ) R_{z} ( \theta _{i} ),
\end{equation}
and the rotation matrices are
$$R_{z}(u)=\left( \begin{array}{ccc}
\cos u & -\sin u & 0 \\
\sin u &  \cos u & 0 \\
0 & 0 & 1
\end{array} \right),$$
$$R_{z}(\theta)=\left( \begin{array}{ccc}
\cos \theta & -\sin \theta & 0 \\
\sin \theta &  \cos \theta & 0 \\
0 & 0 & 1
\end{array} \right),$$
and
$$R_{x}(\phi)=\left( \begin{array}{ccc}
1 & 0 & 0 \\
0 & \cos \phi & -\sin \phi \\
0 & \sin \phi &  \cos \phi
\end{array} \right).$$
It is clear that the magnitude of the vector remains fixed
because all that is ever done is to rotate the original vector
a certain number of times.
The product representation is also
complete but the proof of this is more involved and will not
be included here for brevity (see Ref.~\cite{BD91}). It can be seen,
however, that it exhibits the correct number of degrees of
freedom. A general $N$-harmonic string half in terms of the Fourier
series has $6N+3$ vector coefficients
(Eq.~\ref{fourier}) which must satisfy $4N+1$ nonlinear relations
(Eq.~\ref{constr}) leaving a total of $2N+2$ independent degrees of
freedom (Eq.~\ref{final}). The ranges of the independent
parameters $\theta$ and $\phi$ are polar-like and azimuthal-like
respectively.

\section{The General Odd-Harmonic String}

The magnitude constraint has been solved by generating the
string using the product of rotations. The relation between the centre
of mass terms in the two string halves due to overall spatial
periodicity of closed loops can be solved by taking
an odd-harmonic string with no zeroth harmonic component. It should be
noted  that there are many other harmonic parametrizations with no
zeroth harmonic.  We have chosen the odd-harmonic string because it is
the most simple configuration of this type. With this procedure there
will be a loss of generality --- the loops have an extra inversion
symmetry --- but the problem of finding a set of parameters which
satisfies the relation between the centre of mass terms of any two
string halves in terms of the product representation  resists analytical
solution.

The odd $N$-harmonic string half is given by
\begin{equation}
{\bf c '} _{N} (u)= \rho _{N+2} R_{z} (2u) \rho _{N} R_{z}
(2u) \ldots R_{z} (2u) \rho _{3} R_{z} (u) \rho _{1} {\bf
{\hat z}},
\end{equation}
where $N$ is an odd integer.
Because we want no zeroth harmonic component in the string we
must take the  angle $\phi _{1} = \pm {\pi \over 2} $ in both
halves of the string. This gives a total of four families of strings per
harmonic configuration just as found in Ref. \cite{DES} for the general
three harmonic  string.
In this work we will only analyze those with
$\phi _{1} = {\pi \over 2}$ in both halves of the string.
For an $N/M$ odd-harmonic string (the
first letter denotes the number of harmonics of ${\bf a}$ and
the second that of ${\bf b}$) this leaves us with a total of
$N+M+4$ free parameters ($(N+2)+(M+2)$). Two more parameters, the
$\theta _{1}$ for ${\bf a}'$ and ${\bf b}'$, control the origin of
$u$ and $v$ respectively and are set to $- {\pi \over 2}$ to put
the vector into standard form \cite{BD91}. Because we are only looking
for self-intersections
we can eliminate three extra parameters from overall orientation
freedom, namely $\theta _{N+2}$ from ${\bf a}'$ and $\theta _{N+2}$
and $\phi _{N+2}$ from ${\bf b}'$ leaving $R_{x} (\phi _{N+2})$
in ${\bf a'}$ to ensure
{\it relative} orientation freedom. This leaves a grand total of
$M+N-1$ free parameters for the most general $N/M$ odd harmonic
cosmic loop with no zeroth harmonic. This result agrees with
the number of string parameters of the $3/1$ string presented in
\cite{CDH} and the $3/3$ string presented in \cite{DES}.

When all the redundant parameters have been eliminated the
harmonic expansions in terms of the product of rotations
for the two string halves are
\begin{equation}
{\bf a}' _{N} (u)=  R_{x} (\phi _{N+2}) R_{z} (2u)
\rho_{N}
R_{z} (2u) \ldots R_{z} (2u) \rho _{3} R_{z} (u)  {\bf {\hat x}}
\end{equation}
and
\begin{equation}
{\bf b}' _{M}(v)=  R_{z} (2v) \rho _{M}
R_{z} (2v) \ldots R_{z} (2v) \rho _{3} R_{z} (v)  {\bf {\hat x}}
\end{equation}
These formulae are the expressions for
the derivatives of the string halves. What we need in order
to find self-intersections, however, are the coefficients
of the string halves themselves. Fortunately a recurrence
relation developed in Ref.~\cite{BD91} can be used to compute the
vector coefficients of the Fourier series in terms of the
angles $\theta_{i}$ and $\phi_{i}$. It yields the vector coefficients
in standard form, so the coefficients of $\bf a$ still need to be
multiplied by $R_{x} (\phi_{N+2})$.

We have thus found a means of generating the vector coefficients
of a general $N/M$ odd-harmonic string which satisfies the magnitude
and centre of mass constraints. This is the set of strings which we will
analyze for self-intersections.

\begin{figure}[hbtp]
 \begin{center}
 \begin{tabular}{ccc}
   {\footnotesize(a)} & &
   {\footnotesize(b)}
 \end{tabular}
 \end{center}
\caption{ Two examples of strings generated by the product of rotations
           and the recurrence relation:
          (a) A 19-19 harmonic string and (b) a 47-47 harmonic string.}
 \label{fig:strings}
\end{figure}

The equation to be solved for finding self-intersections is
\begin{equation}
{\bf r} ( \sigma ,t)= {\bf r} ( \sigma ',t)
\end{equation}
which in terms of the Fourier series is
$$\sum  {1 \over {n}} \bigl [ {\bf A }_n \sin(nu) - {\bf B}_n
\cos(nu)+{\bf C }_n \sin(nv)- {\bf D}_n \cos(nv) \bigr ]$$
\begin{equation}
=\sum    {1 \over {n}} \bigl [ {\bf A }_n \sin(nu')- {\bf B}_n
\cos(nu')+{\bf C }_n \sin(nv')- {\bf D}_n \cos(nv') \bigr ]
\label{selfint}
\end{equation}
where
\begin{eqnarray}
u= \sigma -t,&\quad& v= \sigma +t, \\
u '= \sigma  '-t,&\quad& v'= \sigma ' +t,
\label{uvdef}
\end{eqnarray}
and $\sigma \neq \sigma ' + 2n \pi$ for any integer $n$.
This yields three non-linear equations in $\sigma $,
$\sigma '$ and $t$ to be solved simultaneously. Although individual
cases of string parametrizations \cite{KT82,Tur84,CDH,DES,GV87} have been
solved analytically this problem is intractable for a general $N/M$
odd-harmonic string.  Numerically, however, it can be solved. The
general procedure used here is described in the Appendix.  To
eliminate spurious intersections, it is necessary to introduce a
small-distance cutoff.  Self-intersection results are typically
\cite{CDH,DES} given as a function of  the cutoff, and comparisons
between them should only be made for the same cutoffs.

\section{Self-Intersection Results}

The Chen, DiCarlo, Hotes (CDH) string \cite{CDH} has been used to test
the correctness of the self-intersection method developed here.
The percentage of self-intersections for the CDH string has been found
to be $33.6\%\pm 5\%$. This result is in very good agreement with
all the previous work \cite{CDH,DES}. The Turok string \cite{Tur84,CDH}
has also been tested for self-intersections and a percentage of
$2\%\pm 1.25\%$
was found, which is in excellent agreement with the previous work as well.
Furthermore, the $3/3$ string presented here,  which is
equivalent to the DeLaney, Engle and Scheick
(DES) string \cite{DES} has an intersection
probability of about 0.6 (see Figures 3, 4 and 5), this is also consistent
with their results.
The self-intersection probabilities obtained for the generalized
odd-har\-monic strings are presented in the figures below. The
parametrizations for which intersections have been calculated range
from the $3/3$ string to the $25/25$ string for the smallest and largest
cutoffs and from the $3/3$ string to the $41/41$ string for the
middle cutoff.

A random flat parameter distribution has been used for the
angles in the vector coefficients and the number of points $K$ along
the string was chosen to be $K=600$, yielding a
resolution $\eta \approx 0.0104712$ radians which appears to be sufficient.
Three different cutoffs have been used,
$\delta\sigma=|\sigma-\sigma'|=0.084$ radians, $0.126$ radians and
$0.168$ radians, corresponding to eight, twelve and sixteen step lengths,
respectively.  These are approximately the lower, middle and upper
cutoffs  used in the previous work on self-intersections thus allowing
easy comparison.  Each data point on the graphs is an  average value of
ten samples of a hundred strings each.  The errors are the standard
deviation from the average of  these ten samples.

\begin{figure}[hbt]
 \begin{center}
 \begin{tabular}{ccc}
 \end{tabular}
 \end{center}
\caption{Self-intersection probability as a function
of the number of harmonics for a cutoff of
$\delta\sigma=|\sigma-\sigma'|=0.084$ radians.}
\label{fig:inter1}
\end{figure}

\begin{figure}[hbt]
 \begin{center}
 \begin{tabular}{ccc}
 \end{tabular}
 \end{center}
\caption{Self-intersection probability as a function
of the number of harmonics for a cutoff of
$\delta\sigma=|\sigma-\sigma'|=0.126$ radians.}
\label{fig:inter2}
\end{figure}

\begin{figure}[hbt]
  \begin{center}
   \begin{tabular}{ccc}
    \end{tabular}
     \end{center}
      \caption{Self-intersection probability as a function
      of the number of harmonics for a cutoff of
      $\delta\sigma=|\sigma-\sigma'|=0.168$ radians.}
      \label{fig:inter3}
\end{figure}

\begin{figure}[hbt]
 \begin{center}
 \begin{tabular}{ccc}
 \end{tabular}
 \end{center}
\caption{Plot of the logarithm of the stability probability as a function
of the number of harmonics for the three cutoffs. The
strings plotted here are the same as in the previous figures.}
\label{fig:inter4}
\end{figure}

In Figures 3, 4, and 5 plots are shown of the self-intersection
probability as a function of the number of harmonics for the
three cutoffs. It should be noted that the harmonic parametrization
corresponding to a value $N$ on the $x$-axis on the figures corresponds
to a $N/N$ harmonic string. The only parametrizations analyzed
in this work correspond to strings with the {\it same\/} number
of harmonics in each half.

In Figure 6 a plot of the logarithm of the stability
probability as a function of the number of harmonics for the three cutoffs
is presented. The stability
probability is defined as the probability that loop will not self
intersect (i.e. $p_{\rm stab}=1-p_{\rm SI}$). The last six points do not have
errorbars due to the fact that
the lower value of the error is at infinity.  As can be seen the stability
probability $p_{\rm stab}$ can be fitted quite well to an exponential curve.
The dashed curve on the plot is given by
\begin{equation}
 p_{\rm stab} = e ^{\alpha+\beta N}
\end{equation}
with parameters $\alpha = -0.4$ and $\beta = -0.2$. The
continuous curve is
\begin{equation}
 p_{\rm stab} = N^{\delta} e ^{\gamma+\kappa N}
\end{equation}
with parameters $\delta = -0.2$, $\gamma = -0.4$ and
$\kappa = -0.14$,
where $N$ is the number of harmonics.
It is clear that in the high harmonic limit all loops will self-intersect.

\section{Discussion and Conclusions}

We have set out to examine the self-intersection properties
of high-harmonic cosmic loops. After the very early stages of string
formation, damping by the dense surrounding plasma, by the expansion of
the Universe, or by the emission of gravitational radiation does not have
significant effects on the time scale of the order of the oscillation
period.  Consequently, these processes are irrelevant to the question of
how many loops self-intersect.

We have found the most general $N/M$ odd-harmonic string  in terms of the
product of rotations and devised a means for finding self-intersections.
The results agreed with all previous work  on the subject, but extended it
to much higher harmonics.  We have systematically analyzed the resulting
string configurations to find the intersection probability in
the high-harmonic limit.  We have found that
as a function of the number of harmonics the intersection probability
is well described by an exponential.  In the high-harmonic limit almost all
loops self-intersect.

Although loops have been found not to be as important for
structure formation in recent numerical simulations of cosmic string
networks as was once thought, they may still have significant
effects, on the density perturbations \cite{dens}, on the microwave
background \cite{CMBR}, on the gravity-wave background \cite{grav}, and
possibly on the baryon asymmetry \cite{bary}.  On the  basis of our
results, it appears that cosmic loops formed {\it after\/} the initial
string formation era by the mechanisms described in Section 1 will almost
all self-intersect within the first period of oscillation.  As a result of
the intersection process two or more  kinky daughter loops are produced
and, by the same line of reasoning, it is extremely probable that the
daughter loops will re-intersect.  It is not certain how far this process
will go, but it seems likely to continue until the resulting loops have
only a very few kinks, by which time they will be extremely small.
Clearly it would be desirable to follow the evolution of the loops
through several generations to see whether they continue to divide.
Work in this direction is in progress.
Bennett and Bouchet \cite{BB88,BB89} and Allen and Shellard \cite{AS}
found in their simulations that the size of most of their child loops was
determined by the resolution of their simulations, {\it i.e.}, close
enough to the lower cutoff on loop size to suppress the chances of further
division. In this work we have formulated loop trajectories analytically
in an attempt to complement the numerical simulations made so far.  We
believe our method can provide a more definite answer to the question of
the fate of cosmic string loops.

At birth, the loops are already fairly small compared to the horizon.  A
sequence of self-intersection processes will turn the initial parent loop
into tiny loops, with very short lifetimes, so that they will decay into
relativistic particles within a relatively short period of time.  This
is likely to reduce many of the observable effects of loops, in
particular their effects on density perturbations, on the microwave
background and on the gravitational-wave background.  But it might
enhance the effect on the baryon asymmetry.

\appendix

\section*{Appendix}

Here we desribe the implementation of our analysis.

When the number of harmonics for each half of the string
has been selected, a set of string coefficients is randomly generated using
the recurrence relation mentioned in Section 3.  The number $K$ of points
along the string must be chosen.  It determines the resolution $\eta$ of
the string in $\sigma$  and $t$, via $\eta = 2\pi/K$.  For consistency the
resolution in $\sigma$ and $t$ should be the same.  It should be small
enough to ensure that the small scale  detail of the loop is not lost.

The string is then time evolved for the given number of timesteps $K/2$,
with $t$ ranging from $0$ to $\pi$, the effective period.  At each
timestep $t_{i}$, we search for self-intersections by expanding the
equations (\ref{selfint}) to first order about every pair of points $u
_{j}$, $v _{j}$ and  $u _{k}$, $v_{k}$ given by
 \begin{eqnarray*}
u_{j}=\sigma_{j}-t_{i},&\quad& v_{j}=\sigma_{j}+t_{i},\\
u_{k}=\sigma_{k}-t_{i},&\quad& v_{k}=\sigma_{k}+t_{i},
 \end{eqnarray*}
where $j$ ranges\footnote{The symmetry of this set
of strings is such that an intersection found at a point $\sigma _{i}$
on the string will mean that there is another intersection at
$\sigma _{i}+\pi$, so we only need to search for self-intersections
in half the total range in $\sigma$.} from 1 to $K/2$, and $k$ from $j$
to  $K/2$ and from $j+K/2$ to $K$, as follows
 $$
\sum {1\over{n}}\Bigl\{{\bf A}_{n}[\sin(nu_{j})+n\cos(nu_{j})(u-u_{j})]
 -{\bf B}_{n}[\cos(nu_{j})-n\sin(nu_{j})(u-u_{j})]
 $$
 $$
+{\bf C}_{n}[\sin(nv _{j})+n\cos(nv_{j}) (v-v_{j})]-
{\bf D}_{n}[\cos(nv_{j})-n\sin(nv_{j})(v-v_{j})]\Bigr\}
 $$
 $$
=\sum {1\over{n}}\Bigl\{{\bf A}_{n}[\sin(nu _{k})+n\cos(nu_{k})
(u'-u_{k})]-{\bf B}_{n}[\cos(nu_{k})-n\sin(nu_{k})(u'-u_{k})]
 $$
 $$
+{\bf C}_{n}[\sin(nv _{k})+n\cos(nv_{k}) (v'-v_{k})]-
{\bf D}_{n}[\cos(nv_{k})-n\sin(nv_{k})(v'-v_{k})]\Bigr\}.
 $$
Substituting for $u$, $v$, $u'$, and $v'$ from (\ref{uvdef})
and rearranging the terms yields
 $$
\sigma \sum \Bigl\{{\bf A}_{n}\cos(nu_{j})+{\bf B}_{n}\sin(nu _{j})
+{\bf C}_{n}\cos(nv_{j})+{\bf D}_{n}\sin(nv_{j})\Bigr\}
 $$
 $$
-\sigma' \sum \Bigl\{{\bf A}_{n}\cos(nu_{k})+{\bf B}_{n}\sin(nu _{k})
+{\bf C}_{n}\cos(nv_{k})+{\bf D}_{n}\sin(nv_{k})\Bigr\}
 $$
 $$
 +t \sum \Bigl\{-{\bf A}_{n}[\cos(nu_{j})-\cos(nu_{k})]
 -{\bf B}_{n}[\sin(nu_{j})-\sin(nu_{k})]
 $$
 $$
 +{\bf C}_{n}[\cos(nv_{j})-\cos(nv_{k})]
 +{\bf D}_{n}[\sin(nv_{j})-\sin(nv_{k})]\Bigr\}
 $$
 $$
= -\sum \Biggl\{{\bf A}_{n}\left[\left({1 \over n}\sin(nu
 _{j})-u_{i}\cos(nu_{j})\right)-\left({1 \over n}\sin(nu_{k})
 -u_{j}\cos(nu_{k})\right)\right]
$$
$$
 -{\bf B}_{n}\left[\left({1 \over n}\cos(nu
_{j})+u_{j}\sin(nu_{j})\right)
-\left({1 \over n}\cos(nu _{k})+u_{k}\sin(nu_{k})\right)\right]
 $$
$$
+{\bf C}_{n}\left[\left({1 \over n}\sin(nv
     _{j})-v_{j}\cos(nv_{j})\right)
   -\left({1 \over n}\sin(nv _{k})-v_{k}\cos(nv_{k})\right)\right]
 $$
 $$
-{\bf D}_{n}\left[\left({1 \over n}\cos(nv
    _{j})+v_{i}\sin(nv_{j})\right)
  -\left({1 \over n}\cos(nv_{k})+v_{j}\sin(nv_{k})\right)\right]\Biggr\}.
 $$
As can be seen this is a linear system of three equations
and three unknowns $\sigma$, $\sigma'$ and $t$,
which is readily solvable by Cramer's rule. An intersection is
found when the conditions
$|t_{i}-t| < \eta/2$, $|\sigma_{j}-\sigma|< \eta/2$
and $|\sigma_{k}-\sigma'|< \eta/2$
are satisfied. This makes the choice of resolution crucial to
the accuracy of our results. The
process is systematically repeated for every timestep
until an intersection is found.

To avoid trivial solutions ({\it i.e.}, $\sigma = \sigma'$) a cutoff
$\delta\sigma$ has been introduced. This means that if an intersection
occurs at $\sigma_{j}$ and $\sigma_{k}$ with $|\sigma_{j}-\sigma_{k}|$
smaller than the cutoff it is neglected.  In practice, the Taylor
expansion is not made for pairs of points for which
$|\sigma_{j}-\sigma_{k}|<\delta\sigma$.

\end{document}